\newcommand{\be}{\begin{equation}}
\newcommand{\ee}{\end{equation}}
\newcommand{\bea}{\begin{eqnarray}}
\newcommand{\eea}{\end{eqnarray}}
\newcommand{\ba}[1]{\begin{array}{#1}}
\newcommand{\ea}{\end{array}}
\documentclass[12pt]{iopart}
\usepackage{graphicx}
\usepackage{bbm}
\begin{document}
\title{Decoherence effects in interacting qubits under the influence of various environments}
\author{Sumanta Das and G. S. Agarwal }
\address{Department of Physics, Oklahoma State University, Stillwater, Oklahoma 74078, USA}
\eads{\mailto{sumanta.das@okstate.edu},
\mailto{girish.agarwal@okstate.edu}}
\date{\today}
\begin{abstract}
We study competition between the dissipative and coherent effects in the entanglement dynamics of two  qubits. The coherent interactions are needed for designing logic gate operations with systems like ion traps, semicondutor quantum dots and atoms. We show that the interactions lead to a phenomenon of periodic disentanglement and entanglement between the qubits. The disentanglement is primarily caused by environmental perturbations. The qubits are seen to remain disentangled for a finite time before getting entangled again. We find that the phenomenon is generic and occurs for wide variety of models of the environment. We present analytical results for the time dependence of concurrence for all the models. The periodic disentanglement and entanglement behavior is seen to be precursor to the sudden death of entanglement (ESD) and can happen, for environments which do not show ESD for noninteracting qubits. Further we also find that  this phenomenon can even lead to delayed death of entanglement for correlated environments.
\end{abstract}
\maketitle
\section{Introduction}
It is now well understood that entanglement is the key resource for implementation of many quantum information protocols like teleportation, cryptography, logic operations and quantum communications \cite{niel, duetch, shor, bennett, zei, divincenzo}. Bi-partite entanglement i.e entanglement among two quantum mechanical systems each envisaged as a quantum bit (a quantum mechanical two level system analogous to a classical bit), has been found to be particularly important in this context. Numerous methods of producing qubit-qubit entanglement have been investigated during the past decade. A method, which is of particular interest in the context of quantum logic gate operations with systems like ion-traps and semiconductor nanostructures, relies on the coherent interactions among the qubits \cite{zoller, wineland, barenco, vin,li,cal,atac,petta,hans,rob}. An earlier proposal by Barenco \textit{et. al.} \cite{barenco} has shown how one can implement a fundamental quantum gate like the C-NOT gate using dipole-dipole interaction among two quantum dots modeled as two qubits.This was followed by another proposal from DiVincenzo and Loss \cite{vin} in which they showed how the Heisenberg exchange interaction between two quantum dots can be used to implement universal one and two-qubit quantum gates. In their model the qubit is realized as the spin of the excess electron on a single-electron quantum dot. They proposed the electrical gating of the tunneling barrier between neighbouring quantum dots to creat an Heisenberg coupling between the dots. Finally they showed explicitly how by controlling the exchange coupling one can implement a quantum swap gate and XOR operation. Moreover they also showed the implementation of single qubit rotation using pulsed magnetic field. Further in a later work Cirac and Zoller \cite{zoller} discovered that by using the coulombic interaction among two ions one can implement a two-qubit quantum logic gate operation. Clearly many proposals require interacting qubits for two qubit quantum gates.  \\
\indent{}However for a computation to progress efficiently one needs sustained entanglement among the qubits as they dynamically evolve in time. This can be achieved effectively if the quantum mechanical system under evolution is weakly interacting with its surrounding. In practice though as the system evolves the system - environment interaction becomes stronger thereby inhibiting loses in its initial coherence. This loss of quantum coherence is known as decoherence \cite{zurek} and leads to degradation of entanglement. Thus the study of dynamical evolution of two entangled qubits coupled to environmental degrees of freedom is of fundamental importance in quantum information sciences. In recent years numerous studies have been done in this respect \cite{hor, raj, diosi, daf, dod, tin, mint}. One study in particular predicted a remarkable new behavior in the entanglement dynamics of a bi-partite system. It reported that a mixed state of an initially entangled two qubit system, under the influence of a pure dissipative environment becomes completely disentangled in a finite time \cite{tin}. This was termed as \textit{Entanglement Sudden Death} (ESD) \cite{eberly} and was recently observed in two elegantly designed experiments with photonic qubits \cite{almeida} and atomic ensemble \cite{kimble}. Note that an earlier proposal have discussed pausible  experiment to observe ESD in cavity QED and trapped ion systems \cite{sant_pr}. The phenomenon of ESD have motivated numerous theoretical investigation in other bipartite systems involving pairs of atomic, photonic, and spin qubits \cite{mar, gong,tol,chou}, multiple qubits \cite{lopez} and spin chains \cite{cor, lai, abliz}. Further ESD has also been studied for different environments including collective vacuum noise \cite{ficek1}, classical noise \cite{eberly1} and thermal noise \cite{ikram,liu,james}. Moreover random matrix environments have been studied \cite{gorin,pin}. These authors \cite{pin}, also point out the differences 
in the time evolution of concurrence arising from the internal dynamics of two entangled qubits due to the level splitting of each qubit.\\
\indent{}ESD in continuous variable systems has also been extensively studied. In particular the problem of oscillators interacting with different environments has attracted lots of interest \cite{dod, illuminati, paris, prauz, benatti,paz}.  Note that the conditions leading to ESD and probable ways of suppressing it are currently being actively investigated \cite{sant_pr,lidar}. In particular it has been shown how ESD can be avoided by using external modulation with an electromagnetic field \cite{gordon, tahira, paraoanu2} and can even lead to sudden birth for some cases \cite{yu}. Moreover sudden birth of entanglement has also been predicted for structured heat baths \cite{maz, zell} and certain choice of initial conditions of the entangled qubits \cite{ficek}. In another recent work it has been shown that under a pure dephasing environment for a general two mode N-photon state ESD does not occur \cite{asma}. This result was explicitly proven for a general 3-photon state of the form $|\Psi\rangle = a|30\rangle+b|21\rangle+c|12\rangle+d|03\rangle$. \\
\indent{}Even though numerous investigations on ESD in a variety of systems have been done so far, the question of ESD in interacting qubits remains yet open. In this paper we investigate this question for a system of interacting qubits in contact with various models of the environment. We show that due to coherent qubit-qubit interactions two initially entangled qubits, get repeatedly disentangled and entangled as they dynamically evolve leading to dark and bright periods in entanglement \cite{das}.  Moreover we find that the amplitude of bright periods reduce with time and eventually at some finite time vanishes completely, thereby causing death of entanglement. Our investigations also reveal that the length of the dark periods depends on the initial condition of the entangled qubits and also on the interaction strength. Further we find dark and bright periods in entanglement in presence of interaction among the qubits, for initial states which do not exhibit sudden death but simple asymptotic decay of entanglement in absence of the interaction. We find \textit{the existence of dark and bright periods to be generic for interacting qubits and occurs for a wide variety of models for the environment}. We show this explicitly by considering various models of the environment which induce correlated decays, pure and correlated dephasing of the qubits. All of these models exhibit the phenomenon of dark and bright periods even though some of them don't show ESD.\\
\indent{}The organization of this paper is as follows. In section II we discuss the model for two interacting qubits in contact with a simple dissipative environment and formulate their dynamical evolution by solving the quantum-Louiville equation of motion. In section III we develope the theory to study the dynamics of entanglement of the two interacting qubits and calculate the time evolution of the concurrence under the influence of environmental perturbations. In section IV we then study the entanglement dynamics of two interacting qubits under the influence of pure dephasing environment. We find that coherent qubit-qubit interaction not only leads to dark and bright periods in entanglement it also delays the onset of ESD. Further in section V we do a detail study of the dynamics of qubit-qubit entanglement for both non-interacting and interacting qubits for two different correlated models of the environment. In section V A we focus on dissipative environments inducing correlated decay of the qubits. Here we find that for non-interacting qubits there is no ESD and even though entanglement vanishes for certain initial conditions at some instant, it gets partially regenerated quickly and then decays very slowly. When we include the interaction among the qubits we find that entanglement exhibits the phenomenon of dark and bright periods. We further study the behavior of two qubit entanglement for a pure correlated dephasing environment in section V B. We find that the correlated dephasing leads to delay of ESD in absence of qubit-qubit interactions. We see that the degree of delay depends on the strength of the correlation. Here again when we include the qubit-qubit interaction we observe dark and bright periods in entanglement with a much later onset of ESD. In each section we mention the earlier works. Finally in section VI we summarize our findings and conclude with  future outlook.\\

\section{Qubit-Qubit Interaction}
The model that we consider for our study consist of two initially entangled interacting qubits, labeled A and B. Each qubit can be characterized by a  two-level system with an excited state $|e\rangle$ and a ground state $|g\rangle$. Further we assume that the qubits interacts independently with their respective environments. This leads to both local decoherence as well as loss of entanglement of the qubits. The decoherence, for instance can arise due to spontaneous emission from the excited states. Figure 1. show a schematic diagram of our model. 
\begin{figure}[!h]
\begin{center}
\includegraphics[scale = 0.35]{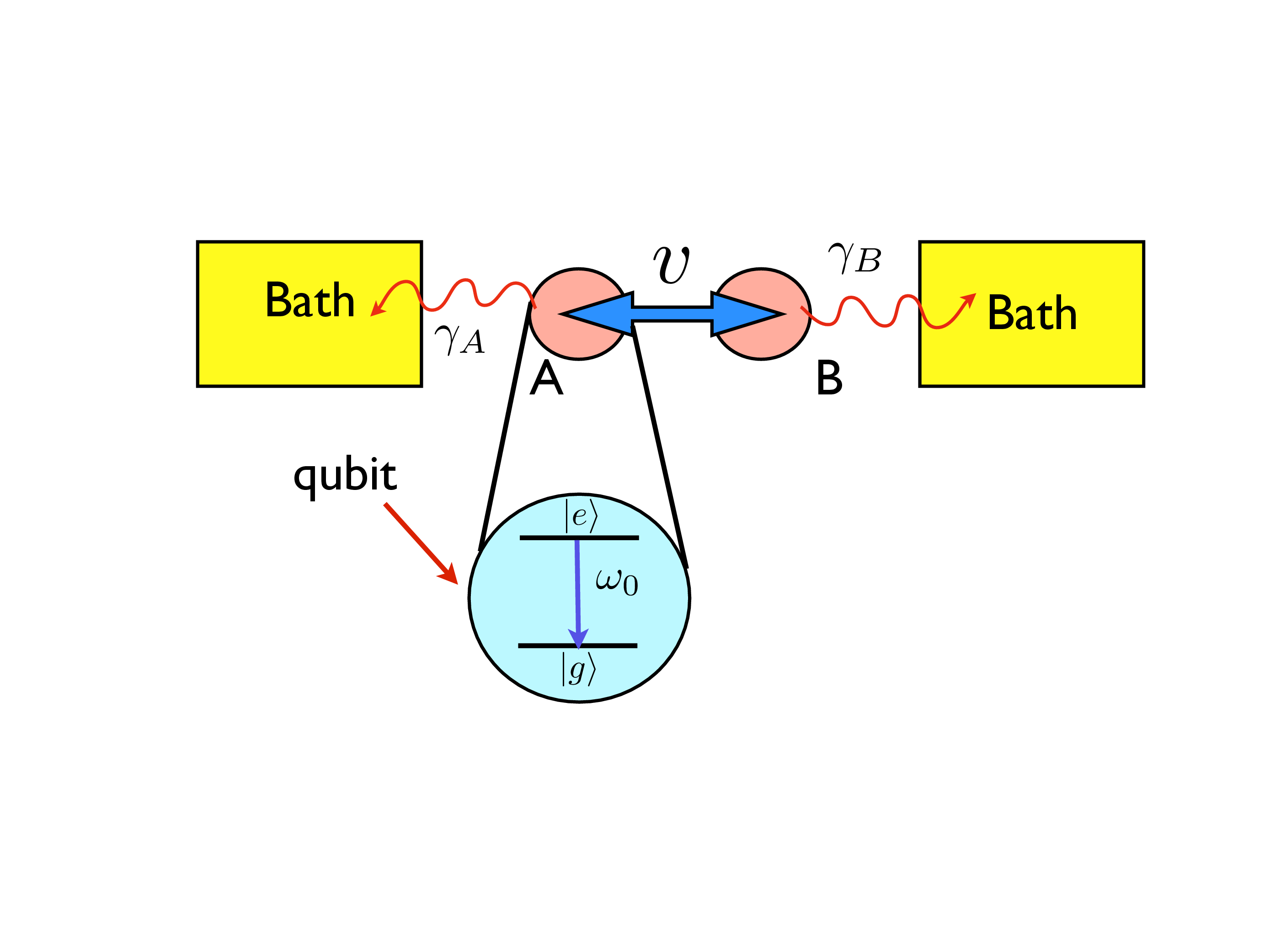}
\caption{(Color online) Schematic diagram of two qubits modelled as two two-level atom coupled to each other by an interaction parameter $v$. Here $|e\rangle, |g\rangle$ signifies the excited and ground states and $\omega_{0}$ their corresponding transition frequency.The qubits A and B independently interact with their respective environments (baths) which lead to local decoherence as well as loss in entanglement.}
\end{center}
\end{figure}
The Hamiltonian for our model is then given by,
\be
\label{1}
\mathcal{H} = \hbar \omega_{0}(S^{z}_{A}+S^{z}_{B})+\hbar v(S^{+}_{A}S^{-}_{B}+S^{+}_{B}S^{-}_{A}),
\ee
where $v$ is the interaction between the two qubits, $S^{z}_{i},S^{+}_{i},S^{-}_{i}$ ($i = $A,B) are the atomic energy, raising and lowering operators defined as $S^{z}_{i} = 1/2(|e_{i}\rangle\langle e_{i}|-|g_{i}\rangle\langle g_{i}|), S^{+}_{i} = |e_{i}\rangle\langle g_{i}| = (S^{-}_{i})^{\dagger}$ respectively and obey angular momentum commutation algebra. We would use the two qubit product basis given by,
\bea
\label{2}
|1\rangle  = |e\rangle_{A}\otimes|e\rangle_{B}&\qquad& |2\rangle  = |e\rangle_{A}\otimes|g\rangle_{B}\nonumber\\
|3\rangle = |g\rangle_{A}\otimes|e\rangle_{B}&\qquad& |4\rangle  = |g\rangle_{A}\otimes|g\rangle_{B}
\eea
Now as each qubit independently interacts with its respective environment, the dynamics of this interaction can be treated in the general framework of master equations. The time evolution of the density operator $\rho$ which gives us information about the dynamics of the system can then be evaluated from the quantum-Liouville equation of motion,
\bea
\label{3}
\dot{\rho}=-\frac{i}{\hbar}[\mathcal{H},\rho]+\mathcal{L}\rho ,
\eea
where  $\mathcal{L}\rho$ includes the effect of interaction of the environment with the qubits. Note that in its simplest form this can be considered to be a spontaneous emission process induced by the vacuum fluctuation of the radiation field. For the case of simple dissipative environment with which the qubits are interacting independently, the effect will be decay of the excited state and any initial coherences of the qubit.  As an example say for qubit A this can be written as,
\bea
\label{3a}
\dot{\rho}_{ee} & =& -2\gamma_{A}\rho_{ee} \nonumber\\
\dot{\rho}_{eg} & = & - \gamma_{A}\rho_{eg}.
\eea
The above equation together with the normalization $\Tr[\rho] =1$ and symmetry of the density matrix, define completely the dynamical system. The effect of environment as elucidated in equation (\ref{3a}) can be written in a compact form in terms of the atomic operators $S^{+},S^{-}$ as,
\be
\label{4}
\mathcal{L}\rho = -\sum_{j = A, B}\frac{\gamma_{j}}{2}(S^{+}_{j}S^{-}_{j}\rho-2S^{-}_{j}\rho S^{+}_{j}+\rho S^{+}_{j}S^{-}_{j}),
\ee
where the terms $\gamma_{A}(\gamma_{B} )$ gives the decay rate of qubit A (B) to the environment.
We give the complete analytical solution of equation (\ref{3}) in the basis defined by (\ref{2}) for coupling to a dissipative environment (\ref{4}) in appendix A. 

\section{Concurrence Dynamics}
To investigate the effect of interaction among the two qubits on decoherence we need to study the dynamics of two qubit entanglement. The entanglement for any bipartite system is best identified by examining the concurrence \cite{wot,buch}, an entanglement measure that relates to the density matrix of the system $\rho$. The concurrence for two qubits is defined as,
\be
\label{5}
C(t) = \max\{0, \sqrt{\lambda_{1}}-\sqrt{\lambda_{2}}-\sqrt{\lambda_{3}}-\sqrt{\lambda_{4}}\},
\ee
where $\lambda$'s are the eigenvalues of the non-hermitian matrix $\rho(t)\tilde{\rho}(t)$ arranged in non-increasing order of magnitude. The matrix $\rho(t)$ being the density matrix for the two qubits and the matrix $\tilde{\rho}(t)$ is defined by,
\be
\label{6}
\tilde{\rho}(t)  = (\sigma^{(1)}_{y}\otimes\sigma^{(2)}_{y})\rho^{\ast}(t)(\sigma^{(1)}_{y}\otimes\sigma^{(2)}_{y}),
\ee
where $\rho^{\ast}(t)$ is the complex conjugation of $\rho(t)$ and $\sigma_{y}$ is the well known time reversal operator for spin half systems in quantum mechanics. 
Note that concurrence varies from $C = 0$ for a separable state to $C = 1$ for a maximally entangled state. Though in general the two qubit density matrix $\rho$ will have all sixteen elements, here we consider the initially entangled qubits to be in a mixed state \cite{tin} given by the density matrix,
\bea
\label{8a}
\rho &\equiv& 1/3(a|1\rangle\langle 1|+d|4\rangle\langle 4|+(b+c)|\psi\rangle\langle\psi|);\nonumber\\
|\psi\rangle & = & \frac{1}{\sqrt{b+c}}(\sqrt{b}|2\rangle+e^{i\chi}\sqrt{c}|3\rangle);\nonumber\\
& &\frac{a+b+c+d}{3}  =  1;
\eea
where $a,b,c$ are independent parameters governing the nature of the initial state of the two entangled qubits. Note that the entanglement part of the state depends on the initial phase $\chi$. Following (\ref{8a}) one can see that the initial two qubit density matrix have only six-elements. In the matrix form $\rho$ is then given by, 
\bea
\label{9}
\rho(0) = \frac{1}{3}
\left(\begin{array}{cccc} a & 0 & 0 & 0\\
 0 & b & z & 0\\
 0 & z^{\ast} & c & 0 \\
 0 & 0 & 0 & d\ \end{array}\right).
\eea
Here $z = e^{i\chi}\sqrt{bc}$ are the single photon coherences. Using the solution of the quantum-Liouville equation (\ref{4a}) it can be shown that the \textit{initial density matrix} (\ref{9}) \textit{preserves its form for all t}. Finally we calculate the concurrence defined by (\ref{5}) and (\ref{6}) for the two qubits as,
\be
\label{10}
C(t) = \mathsf{Max}\lbrace 0, \tilde{C}(t)\rbrace,
\ee
where $\tilde{C}(t)$ is given by,
\be
\label{11}
\tilde{C}(t)  = 2\left\lbrace|\rho_{23}(t)|-\sqrt{\rho_{11}(t)\rho_{44}(t)}\right\rbrace
\ee
Let us now consider a particular class of mixed states with a single parameter $a$ satisfying intially $a \geq 0$, $b = c = |z| = 1$ and $d = 1-a$ \cite{tin}. Note that then (\ref{8a}), has the structure similar to a Werner state \cite{wer}. On using the dynamical evolution of the density matrix elements from appendix (A) and this set of initial conditions in (\ref{11}), we obtain,
\bea
\label{12}
\tilde{C}(t) & = &\frac{2}{3}e^{-\gamma t}\lbrack(\cos^{2}\chi+\sin^{2}\chi\cos^{2}(2vt))^{1/2}\nonumber\\
&-&\sqrt{a(1-a+2w^{2}+w^{4}a)}\rbrack,
\eea
\begin{figure}[!h]
\vspace{0.3in}
\begin{center}
\includegraphics[scale = 0.55]{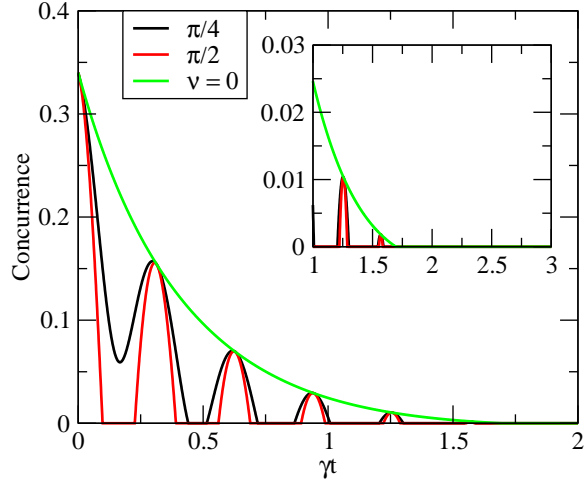}
\caption{(Color online) Concurrence as a function of time for two initially entangled, interacting qubits with initial conditions $b = c = |z| = 1.0 $ and two different initial phases $\chi = \pi/4$ (black curve) and $\chi = \pi/2$ (red curve).}
\end{center}
\end{figure}
where $w = \sqrt{1-e^{-\gamma t}}$. One can clearly see the dependence of $\tilde{C}(t)$ on the interaction $v$ among the qubits and the initial phase $\chi$. We see from (\ref{12}) that in absence of the interaction $v$, concurrence becomes independent of the initial phase and yields the well established result of Yu and Eberly \cite{tin}.\\
Note that $\tilde{C}(t)$ can become negative if,
\bea
\label{13}
 a(1-a+2w^{2}+w^{4}a) > (1-\sin^{2}\chi\sin^{2}(2vt)),
 \eea
in which case concurrence is zero and the qubits get disentangled. In figure (2) we show the time dependence of the entanglement by plotting equation (\ref{12}) for $v = 5\gamma$ , $a = 0.4$ and  different values of the initial phase $\chi$. 
\begin{figure}
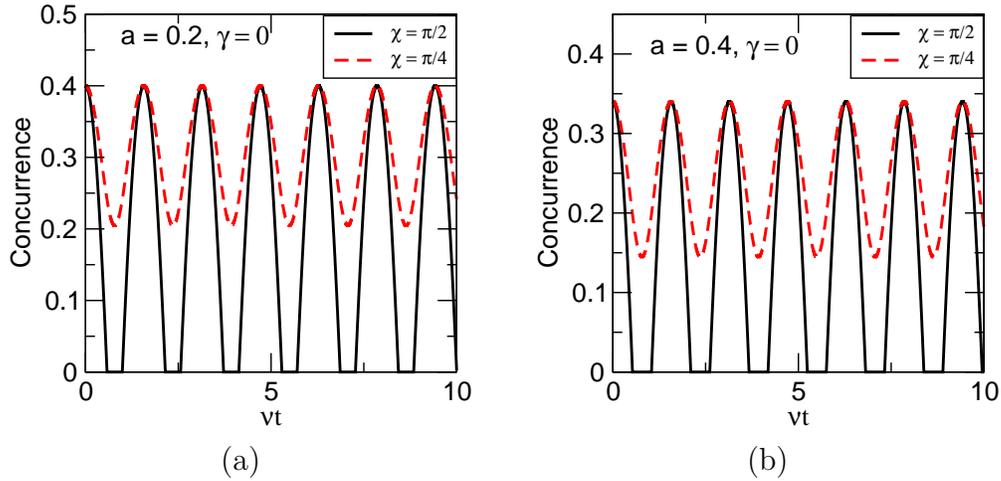

\vspace{0.2in}
\begin{center}
\begin{tabular}{ccccc}
\includegraphics[scale = 0.42]{p3.eps} & & \includegraphics[scale= 0.42]{p4.eps}\\
(a) & & (b)
\end{tabular}
\caption{(Color online) Evolution of Concurrence for two initially entangled, interacting qubits with initial conditions $a = 0.4, b = c = |z| = 1.0 $ and different initial phases $\chi$. Here $\gamma = 0$. The magnitude of bright periods in absence of environment does not diminish in magnitude. }
\end{center}
\end{figure}
The inset of figure 2 shows the long time behavior of entanglement for this case. We see from figure 2 that non-interacting qubits ($v/\gamma = 0$) exhibit sudden death of entanglement (ESD) [visible more clearly in the inset]  whereas when they interact ($v/\gamma \neq 0$) the concurrence oscillates between zero and non-zero values. Thus we see that the initially entangled qubits in presence of interaction $v$ gets repeatedly disentangled and entangled leading to \textit{dark and bright} periods in the concurrence. The magnitude of bright periods diminish with time and eventually at longer time this behavior vanishes completely leading to death of entanglement (ESD). The length of a dark period is determined by the condition (\ref{13}).  We have found that this behavior in entanglement prevails for other values of the parameter $a$ also \cite{das}. In figure 3(a) and (b) we plot the dyamical evolution of entanglement when ($\gamma = 0$) , i.e in absence of any environmental perturbation. This is a ideal case of close quantum systems whose dynamics is only influenced by the initial condition of the entangled qubits and the inter-qubit interactions. In this case we get,
\be
\tilde{C}(t) = \frac{2}{3}[\sqrt{\cos^{2}\chi+\sin^{2}\chi\cos(2vt)}-\sqrt{a(1-a)}]
\ee
For both the case of $a = 0.2$ and $a = 0.4$ with an initial phase of $\chi/4$ , we observe sinusoidal behavior of entanglement as seen in both (a) and (b) of figure 4. Thus there is no ESD in absence of the environment in this case. For another value of the initial phase $\chi/2$ we observe dark and bright periods of entanglement. The periods of disentanglement (dark periods) are governed by the condition $\sqrt{a(1-a)} > |\cos(2vt)|$. It is clearly visible from the plots that in absence of any environment the amplitude of the bright periods does not diminish at all and thus the qubits gets back their initial entanglement completely. This regeneration of entanglement is due to the inter-qubit interactions. Note that similar behavior in concurrence dynamics (collapse and revival of entanglement) have been predicted in earlier studies of non-interacting qubits in atom cavity systems. For example it was shown that for double Jaynes-Cumming (JC) \cite{jaynes_c} model, with completely undamped non-interacting cavities entanglement shows a periodic death and re-birth feature \cite{eberly_jb}. This was attributed to exchange of information between the finite number of cavity modes and the atoms - a new kind of temporary decoherence mechanism. In another work pairwise concurrence was calculated among four qubits, where the qubits were formed by the cavity modes and atoms \cite{eberly_jb1}. Here again JC like interaction between the atom and cavity gives rise to dark and bright period in the entanglement dynamics of the qubits. It was shown that during the period when the concurrence between the cavities vanish, the concurrence between the atoms reaches its peak and vice-versa. This only happens as the cavities where assumed to be lossless with finite number of mode and thus without  environmental decoherence. Further it was shown that for qubits remotely located and in contact with their respective environment when driven independently by single mode quantized field, one gets dark and bright periods of entanglement instead of ESD, a feature similar to single atom behavior in cavity quantum electrodynamics \cite{luis,eberly_ol}. \textit{These works} \cite{jaynes_c, eberly_jb, eberly_jb1, luis, eberly_ol} \textit{differ from our's as we focus on the effect of interaction among the qubits in presence of a decohering environment}. Note that in a more recent work it was shown how oscillators interacting with a correlated finite temperature Markovian bath can lead to dark and bright periods in entanglement for certain initial conditions \cite{paz}. 
\section{Pure Dephasing of the Qubits}
\begin{figure}[!h]
\begin{center}
\includegraphics[scale = 0.4]{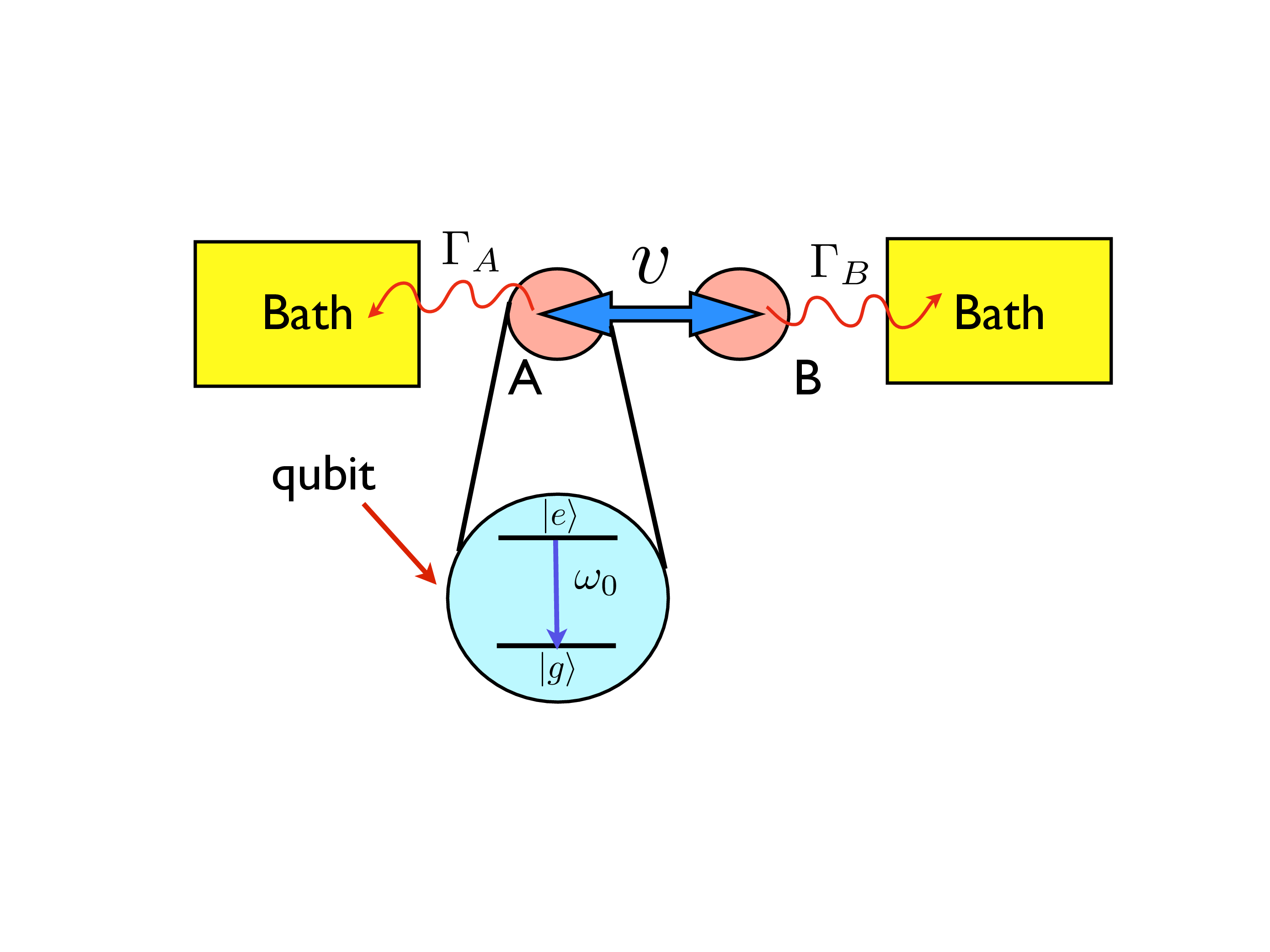}
\caption{(Color online) Schematic diagram of two qubits modelled as two two-level atom coupled to each other by an interaction parameter $v$. Here $|e\rangle, |g\rangle$ signifies the excited and ground states and $\omega_{0}$ their corresponding transition frequency. The qubits A and B independently dephase to their respective environments (baths) which leads to decoherence and thus loss in entanglement. The corresponding dephasing rates are given by $\Gamma_{A}$ and $\Gamma_{B}$ respectively.}
\end{center}
\end{figure}
In order to demonstrate the generic nature of our results, we consider other models of the environment. A model which has been successfully used in experiments \cite{kwait} involves pure dephasing. The mathematical formulation for this kind of an environmental model can be done via a master equation technique and is given by,
\bea
\label{14}
\mathcal{L}\rho = - \sum_{i = A,B}\Gamma_{i}(S^{z}_{i}S^{z}_{i}\rho-2S^{z}_{i}\rho S^{z}_{i}+\rho S^{z}_{i}S^{z}_{i})
\eea
where $\Gamma_{A} (\Gamma_{B})$ is the dephasing rate of qubit A (B). Substituting (\ref{14}) in (\ref{3}) we get the equation for dynamical evolution of the qubits under the influence of this kind of an environment. Note that in this model the populations do not decay as a result of the interaction with the environment whereas the coherences  like $\rho_{23}(t)$ decay as $\rho_{23}(0)e^{-(\Gamma_{A}+\Gamma_{B})t}$. 
\begin{figure}
\begin{center}
\includegraphics[scale = 0.45]{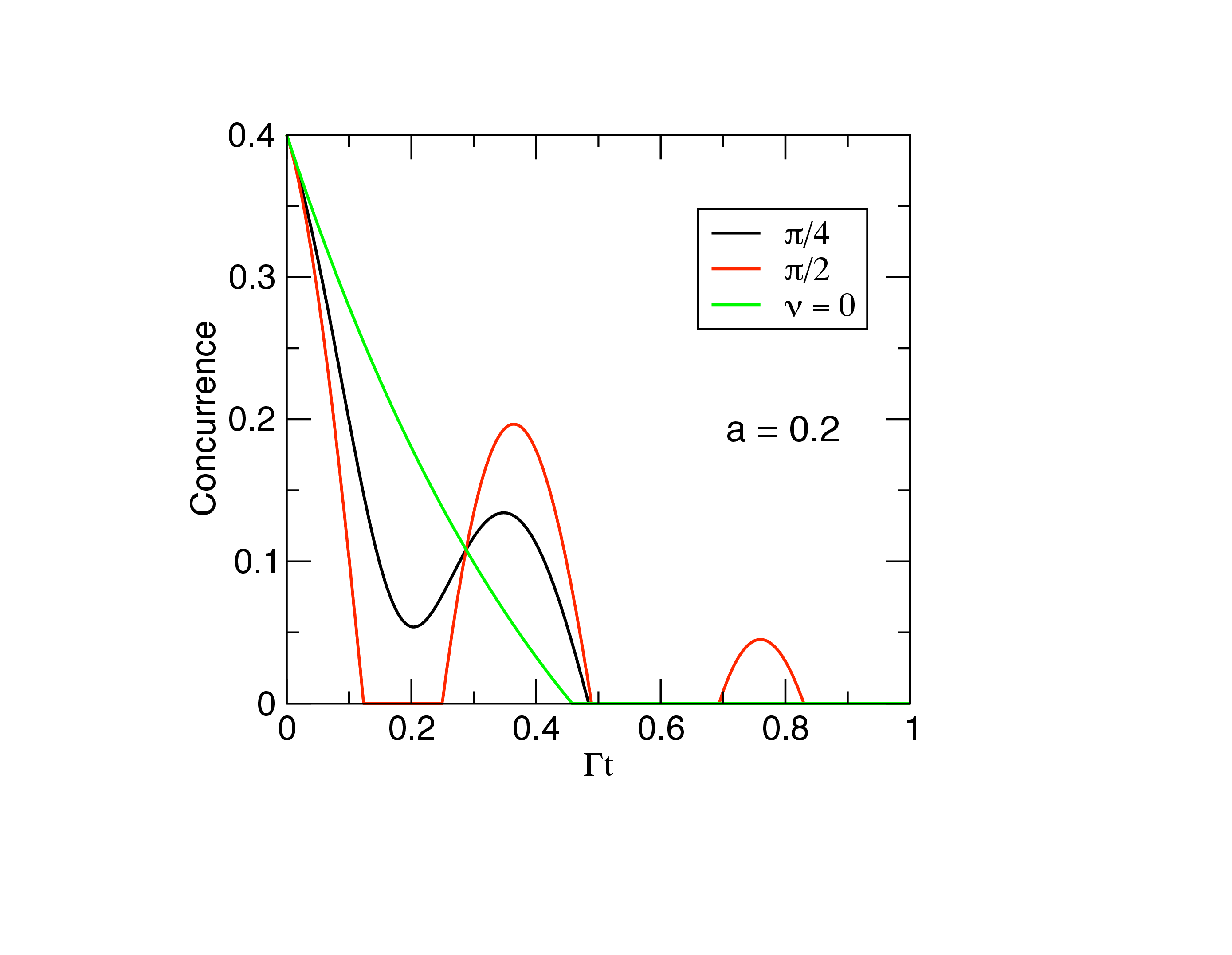}\\
\caption{(Color online) Concurrence as a function of time with initial conditions $ b = c = |z| =1$ and two different values of the initial phase $\chi$ for the dephasing model. The red and black curve in figure is for $\chi = \pi/4$ and $\pi/2$ respectively. Here the interaction parameter is taken to be $v/\Gamma = 4$.}
\end{center}
\end{figure}
Let us now study the the effect of interaction $v$ between the qubits on the dynamics of entanglement. We assume the same initial density matrix of equation (\ref{9}) with the initial conditions $ d = 1-a, b= c = |z| = 1$ and $a \geq 0$ to calculate the concurrence. One can clearly see from the solution of quantum-Louiville equation given in appendix (B) that under pure dephasing, the \textit{form of matrix in} (\ref{9}) \textit{is preserved for all time}.  Using the solutions of the master equation (\ref{3}) derived in appendix (B) for the environment effects given by (\ref{14}) and substituting in equations (\ref{10}), (\ref{11})  we get the time dependent concurrence for this model to be,
\bea
\label{15}
\tilde{C}_{D}(t)& = &\frac{2}{3}\lbrack e^{-\tau}\lbrace e^{-2\tau}\cos^{2}\chi+\sin^{2}\chi\lbrace\cos(\Omega_{1}\tau)\nonumber\\
& &-\frac{1}{\Omega_{1}}\sin(\Omega_{1}\tau)\rbrace^{2}\rbrace^{1/2}-\sqrt{a(1-a)}\rbrack,
\eea
where the suffix $D$ signifies that the concurrence is calculated for a dephasing environment and we assume $\Gamma_{A} = \Gamma_{B} = \Gamma$. Here $\tau = \Gamma t$ and $\Omega_{1} = \sqrt{(2v/\Gamma)^{2} -1}$.
For $v = 0$ we get $\tilde{C}_{D}(t) = 2/3\lbrack e^{-2\tau} -\sqrt{a(1-a)}\rbrack$, which is independent of the initial phase $\chi$. We find \textit{death of entanglement} for $\tau > (1/2)\ln [1/\sqrt{a(1-a)}]$. Note that Yu and Eberly \cite{eberly} have considered this case earlier but for $a = 1$ only, in which case there is no ESD. In figure (5) we show the time dependence of entanglement for a purely dephasing model, for $a = 0.2$ and initial coherences governed by the phase $\chi$.  From the figure we see that  for $v \neq 0$, the two qubit entanglement exhibits the phenomenon of dark and bright periods. Further we also see that for $v \neq 0$, dark and bright periods continues beyond the time when ESD occurs for noninteracting qubits. This kind of behavior in the entanglement dynamics is found for other values of the parameter $a$ and interested readers are refered to \cite{das} for further dicsussions on these.
\section{Concurrence Dynamics in Correlated Environmental Models}
\subsection{Effect of Correlated Dissipative Environment}
\indent{}We next consider an environment involving correlated decay and show how coupling to such environment can lead to new effects in the entanglement dynamics for two qubit systems. We will consider the case of both non-interacting as well as interacting qubits for this model of the environment. To keep the analysis simple and get a better physical insight on the decoherence effect of this environment we will first study the case of non-interacting qubits. We assume as before that the qubits interacts independently with their respective environments with decay rates $\gamma_{A}$ and $\gamma_{B}$.  Further we assume that the qubits are close enough ($r << \lambda$, $r$ being the inter-qubit distance and $\lambda$ the wavelength of emitted radiation in process of a decay) such that they can undergo a correlated decay with decay rates $\Gamma_{AB} (\Gamma_{BA})$ for qubits $A (B)$. Whether this would lead to further decoherence is a question we want to investigate. Note that the entanglement dynamics of two non-interacting two level atoms in presence of dissipation caused by spontaneous emission was studied earlier in details by Jak\'{o}bczyk and Jamr\'{o}z \cite{jj}. They even considered correlated model of dissipative environment and showed possible destruction of initial entanglement and possible creation of a transient entanglement between the atoms.  Further they also discussed the question of non-locality and how it is influenced by the spontaneous emission by explicitly    showing the violation of Bell-CSHS inequality. One of the chief difference between this work and ours is the initial density matrix $\rho$ considered and the interaction introduced between the qubits . While we consider the possibility of both the qubits (atoms) to be initially excited and show its important consequences on the decay dynamics, they have neglected this effect by putting $\rho_{11} (0) = 0$. We would show later in this paper (as can also be seen from their results) that the dissipative environment preserve the form of the initial $\rho$. Hence $\rho_{11} (t) = 0$ for all time in their case.  Moreover, in a recent work the entanglement dynamics of two initially entangled qubits for collective decay model was studied in context to ESD, by Ficek and Tanas \cite{ficek1}. They considered an initial density matrix of the from,
\bea
\rho & = &|\Psi_{0}\rangle\langle\Psi_{0}| ;\nonumber\\
|\Psi_{0}\rangle & =&  \sqrt{p}|e_{1},e_{2}\rangle + \sqrt{(1-p)}|g_{1},g_{2}\rangle
\eea
It can be clearly seen that in this case the two-qubits are initially prepared in an entangled state by the two-photon coherences. They further show that for this initial condition the single photon coherences are never generated. Moreover the dipole-dipole interaction that they consider for the two qubit system have no influence for this initial condition. Ficek and Tanas predicted dark periods and revival in the two qubit entanglement in their work due to the correlated nature of the bath, we on the other hand consider the initial density matrix of the form (\ref{9}) with single photon coherences and show that any coherent interaction among the qubits does influence the entanglement dynamics at all later time. \\ 
We now include the effect of a dissipative environment with both independent and correlated decay of the qubits via a master equation technique given by,  
\bea
\label{16}
\mathcal{L}\rho & =&  - \sum_{j,k = A, B}\frac{\Gamma_{ij}}{2}(S^{+}_{j}S^{-}_{k}\rho-2S^{-}_{k}\rho S^{+}_{j}+\rho S^{+}_{j}S^{-}_{k}), \nonumber\\
\Gamma_{jj} & = & \gamma_{j}
\eea
The time evolution of the density operator $\rho$ which gives us information about the dynamics of the system can then be evaluated by solving the quantum-Liouville equation (\ref{3}) with the environmental effect included by equation (\ref{16}) and taking $v = 0$. Next as before we consider the qubits to be intially entangled with their initial state to be a mixed state defined by the density matrix (\ref{9}). We then solve the quantum-Louiville equation to study the dynamical evolution of the system. The reader is refered to appendix (C) for explicit solution of the time dependent density matrix elements. One can clearly see from appendix (C) that for this kind of model of the environment, as before \textit{the initial density matrix preserves its form for all time t}. Now using appendix (C) in equations (\ref{10}) and (\ref{11}) and the initial conditions $ a \geq 0, d = 1-a$, $b = c = |z| = 1$, we obtain the concurrence dynamics of two initially entangled non-interacting  qubits for this model of the environment as,
\bea
\label{24}
\tilde{C}(t) & = &\frac{2}{3}e^{-\gamma t}\{\lbrack\{\cos\chi\cosh(\Gamma t)-\sinh(\Gamma t)+a\zeta(t)\}^{2}\nonumber\\
&  &+\sin^{2}\chi\rbrack^{1/2}-\sqrt{3a\lbrack1-\kappa(t)\rbrack}\},\nonumber\\
\eea
where $\zeta(t)$ and $\kappa(t)$ are given by , 
\bea
\label{25}
\zeta(t) & = & e^{-\gamma t}\{ \left (\frac{1+\Gamma/\gamma}{1-\Gamma/\gamma}\right )(e^{(1-\Gamma/\gamma)\gamma t}-1)\nonumber\\
& & - \left (\frac{1-\Gamma/\gamma}{1+\Gamma/\gamma}\right )(e^{(1+\Gamma/\gamma)\gamma t}-1) \},
\eea
\bea
\label{26}
\kappa(t)& = & \frac{1}{3}a e^{-2\gamma t}\{ 1+\left (\frac{1+\Gamma/\gamma}{1-\Gamma/\gamma}\right )(e^{(1-\Gamma/\gamma)\gamma t}-1)\nonumber\\
&+ &\left (\frac{1-\Gamma/\gamma}{1+\Gamma/\gamma}\right )(e^{(1+\Gamma/\gamma)\gamma t}-1)\}+ \frac{2}{3}e^{-\gamma t}\{\cosh(\Gamma t)\nonumber\\
& & - \cos\chi\sinh(\Gamma t)\},
\eea
For simplicity we have assumed equal decay rates of both the qubits, $\gamma_{A} = \gamma_{B} = \gamma$ and $\Gamma_{AB} = \Gamma_{BA} = \Gamma$. One can clearly see the dependence of $\tilde{C}(t)$ on the correlated environmental effect given by $\Gamma$ and the initial phase $\chi$ in equation (\ref{24}). We see from (\ref{24}), (\ref{25}) and (\ref{26}) that for $\Gamma = 0$, concurrence becomes independent of the initial phase and yields the result of Yu and Eberly \cite{tin}. Note that $\tilde{C}(t)$ can become negative if,
\bea
\label{27}
3a\lbrack1-\kappa(t)\rbrack  & > & \lbrack\{\cos\chi\cosh(\Gamma t)-\sinh(\Gamma t)+a\zeta(t)\}^{2}\nonumber\\
&  &+\sin^{2}\chi\rbrack
 \eea
in which case concurrence is zero and the qubits get disentangled. 
\begin{figure}
\begin{center}
\includegraphics[scale = 0.43]{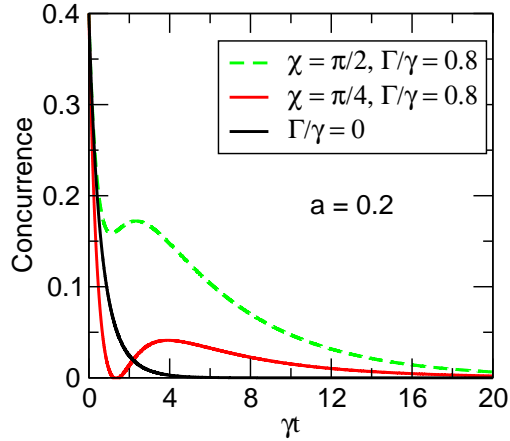}
\caption{(Color online) Time evolution of concurrence for $a = 0.2, b = c =|z| = 1$ and two different initial phases $\chi$ for two non-interacting qubits in contact with a correlated dissipative environment. Here $\Gamma/\gamma = 0$ signifies absence of common bath for the qubits. }
\end{center}
\end{figure}
\begin{figure}
\vspace{0.25in}
\begin{center}
\includegraphics[scale = 0.43]{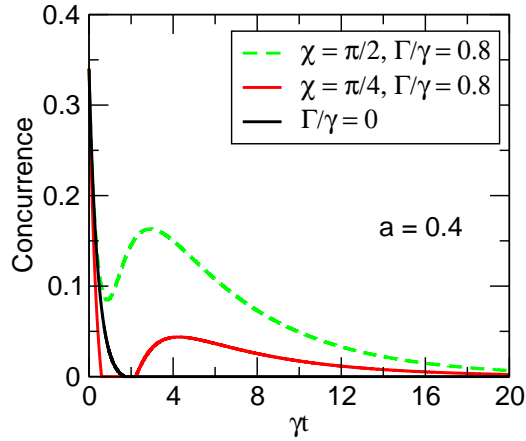}
\caption{(Color online) Time evolution of concurrence govern by the initial condition $a = 0.4$ for two non-interacting qubits in contact with a correlated dissipative environment. Here all other initial parameters remains the same as fig (7). $\Gamma/\gamma = 0$ signifies absence of any common bath in which case entanglement sudden death (ESD) is observed. }
\end{center}
\end{figure}
To understand how correlated decay of the qubits might effect their entanglement we study the analytical 
result of equation (\ref{24}) for different values of the parameter $a$ and $\chi$. In figure (6) we show the time dependence of entanglement for $a = 0.2$ and two different values of initial phase $\chi$ and correlated decay rate of $\Gamma = 0.8\gamma$. Note that for $\Gamma = 0$, there is no ESD in this case \cite{tin} and concurrence monotonically goes to zero as  $t \longrightarrow \infty$. For $\Gamma \neq 0$ we observe new behavior in the entanglement of the qubits. Concurrence is seen to have a much slower decay in comparison to when $\Gamma = 0$. For a initial phase of $\chi = \pi/4$ we observe that the condition in equation (\ref{27}) is satisfied and entanglement vanishes temporarily \textit{i.e} the qubits get disentangled. The entanglement gets regenerated at some later time and finally goes to zero very slowly as $t \longrightarrow \infty$. Note that this disentanglement and re-entanglement phenomenon is non periodic and is very sensitive to initial coherence among the qubits, for example it do not occur when the initial coherence is governed by the phase $\chi = \pi/2$.  In figure (7) we plot concurrence for $a = 0.4$. For this value of $a$ ESD is observed for $\Gamma = 0$ but not for $\Gamma \neq 0$. Instead we observe disentanglement and regeneration of entanglement among the qubits for $\chi = \pi/4$. Here again we find that no dark and bright periods nor any ESD for initial phase of $\chi = \pi/2$. Further, note that for initial phase $\chi = \pi/4$ we have a longer time interval during which the qubits remain disentangled before getting entangled again,  in comparison to the case for $a = 0.2$. Thus we find that the time interval between disentanglement and regeneration of entanglement as well as the magnitude of regeneration strongly depends on the initial coherences of the initially entangled qubits. Hence we can conclude that for non-interacting qubits in contact with a dissipative correlated environment no ESD occurs.\\ 
\indent{}Let us now consider the case of two initially entangled interacting qubits in contact with the correlated environment. The dynamical evolution of the system in presence of interaction $v$ for correlated model of environment is evaluate in details in appendix (D). We use the solutions of appendix (D) in (\ref{11}) to calculate the concurrence for this environment. Note that the solutions are essentially valid under the assumption that our initial two qubit density matrix $\rho$ is given by equation (\ref{9}). Further we consider as before that the two entangled qubit's evolution is governed by the initial conditions $a \geq 0, d =1-a$, $b = c = |z| = 1$. Hence the time dependent concurrence for two initially entangled interacting qubits becomes,
\bea
\label{27a}
\tilde{C}(t) & = &\frac{2}{3}e^{-\gamma t}\{\lbrack\{\cos\chi\cosh(\Gamma t)-\sinh(\Gamma t)+a\zeta(t)\}^{2}\nonumber\\
&  &+\cos^{2}(2vt)\sin^{2}\chi\rbrack^{1/2}-\sqrt{3a\lbrack1-\kappa(t)\rbrack}\},\nonumber\\
\eea
\be
C(t)  = \mathsf{Max}\lbrace 0, \tilde{C}(t)\rbrace\quad;
\ee
where $\zeta(t)$ and $\kappa(t)$ are given by equations (\ref{25}) and (\ref{26}) respectively.  The dependence of concurrence for $\tilde{C}(t) > 0$ on the interaction strength $v$ between the qubits is clearly visible in equation (\ref{27a}). Further now we can see that the condition of complete disentanglement of the qubits is given by,
\bea
\label{27b}
3a\lbrack1-\kappa(t)\rbrack  & > & \lbrack\{\cos\chi\cosh(\Gamma t)-\sinh(\Gamma t)+a\zeta(t)\}^{2}\nonumber\\
&  &+\cos^{2}(2vt)\sin^{2}\chi\rbrack
\eea
\begin{figure}
\begin{center}
\includegraphics[scale = 0.47]{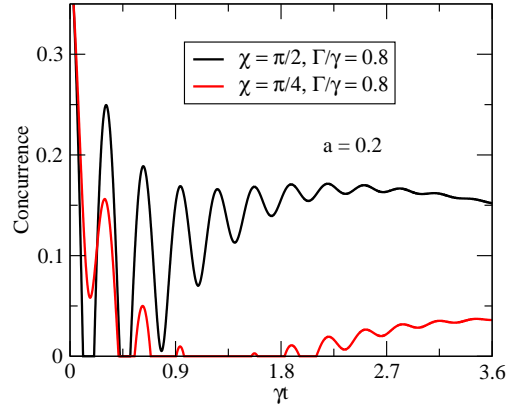}
\caption{(Color online) Time evolution of concurrence for interacting qubits in contact with a correlated dissipative environment with correlated decay rate of $\Gamma/\gamma = 0.8$. Here $b = c = |z| = 1$. A long period of disentanglement is observed for initial phase $\chi = \pi/4$. Here the interaction strength among the qubits is taken to be $v/\gamma = 5.0$}
\end{center}
\end{figure}
\begin{figure}
\vspace{0.3in}
\begin{center}
\includegraphics[scale = 0.47]{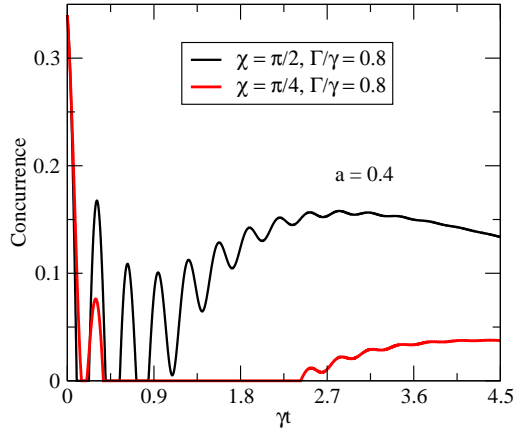}
\caption{(Color online) Time evolution of concurrence for interacting qubits in contact with a correlated dissipative environment for same parameters as figure (9) and $a = 0.4$. The dark and bright periodic features sustain for a longer time for initial phase of $\chi = \pi/2$. Much longer period of disentanglement now observed for $\chi = \pi/4$.}
\end{center}
\end{figure}
When condition (\ref{27b}) is satisfied, $\tilde{C}(t) < 0$ and hence $C(t) = 0$. Next to study the effect of qubit-qubit interaction on the entanglement dynamics we plot the time dependent concurrence for different value of $a$, initial phase $\chi$ and correlated decay rates of $\Gamma = 0.8\gamma $ in figures (8) and (9). We observe in the figures that for an initial phase of $\chi = \pi/2$ concurrence exhibits dark and bright periods at initial time for both $a = 0.2$ and $a= 0.4$. For longer time the concurrence shows a damped oscillatory behavior. We attribute this effect to the competition between the fast inter-qubit interactions $v$ and the environmental decays. For longer time the correlated decay becomes dominant and leads to a slow damped oscillatory decay of the entanglement. For $\chi = \pi/4$ the dark and bright periods are not very pronounced and is over shadowed very quickly by the correlated decay. Note that for this value of initial phase we find that there exist a long period of time during which the qubits remain disentangled. At a much later time entanglement gets regenerated and increases initially and then starts decaying very slowly after . This behavior is quite different from the dark and bright periods seen for other models of the environment. Thus we see that for interacting qubits there is no ESD for this model of the environment. Instead we find dark and bright periods with long period of disentanglement whose occurrence depends on the initial coherence.
\subsection{Delay of ESD by Correlated Dephasing Environment}
\begin{figure}[!h]
\begin{center}
\includegraphics[scale =0.38]{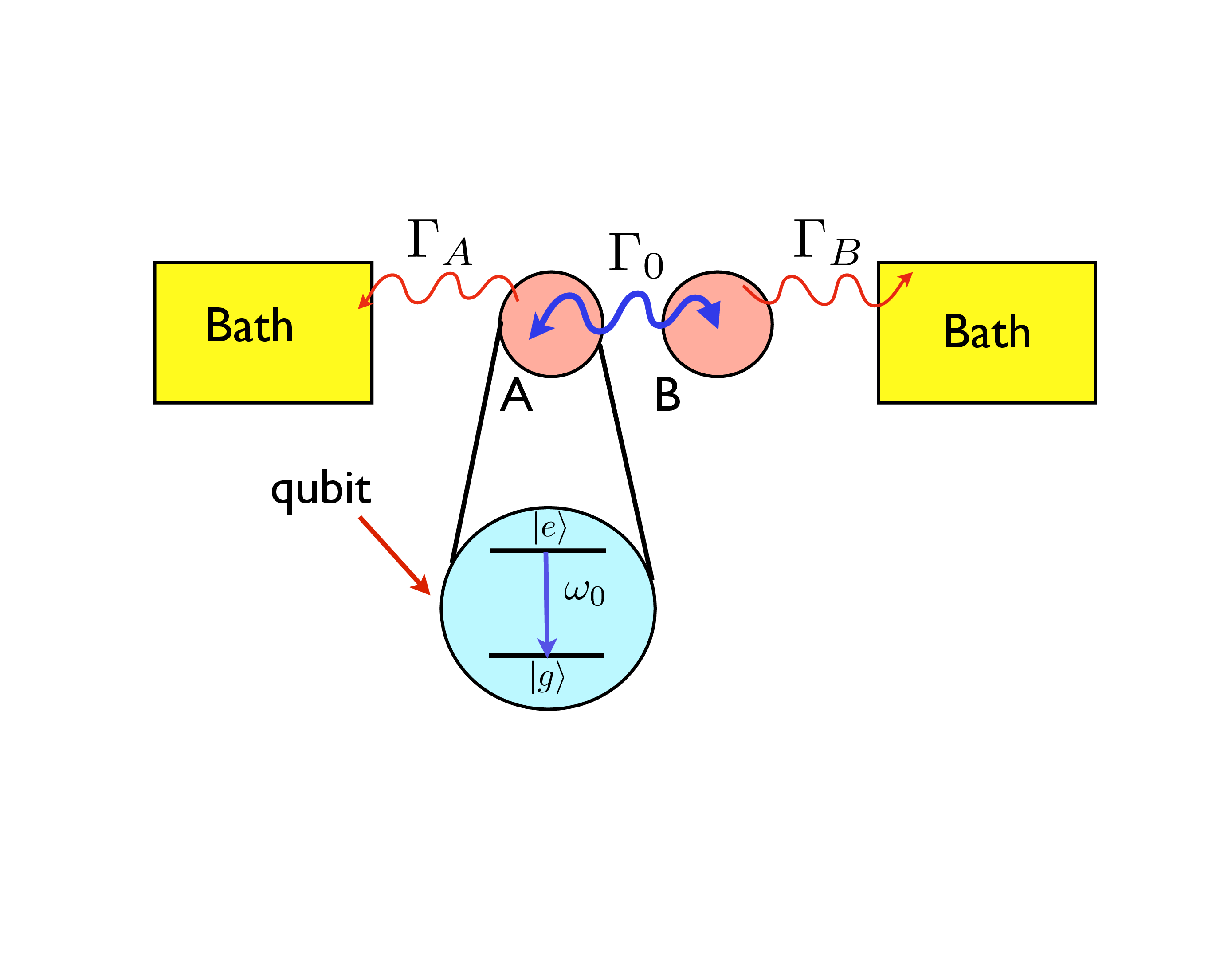}
\caption{(Color online) Schematic diagram of two qubits modelled as two two-level atoms. Here $\omega_0$ is the transition frequency of the excited state $|e\rangle$ to the ground state $|g\rangle$. The qubits A and B independently dephase to their environments (baths) with a dephasing rate of $\Gamma_{A}, \Gamma_{B}$ respectively. The qubits can also interact with the environment collectively when they are at proximity giving rise to correlated dephasing represented by the decay rate $\Gamma_{0}$.}
\end{center}
\end{figure}
\indent{}Finally we consider a purely correlated dephasing model of the environment and study the effect of such an environment on the entanglement dynamics of two qubits. Note that this kind of model is popular among solid state systems like semiconductor quantum dots. We will study the behavior of entanglement for both non-interacting and interacting qubits. As before to keep our analysis simple and to get a better physical insight to the question of decoherence for this kind of environment we will first study the case of non-interacting qubits. We will then generalize our results by introducing the interaction among the qubits. For non-interacting qubits the Hamiltonian for our model is given by (\ref{1}) with $v =0$. The effect of the dephasing environment on the qubits is included via a master equation technique and is given by, 
\bea
\label{28}
\mathcal{L}\rho & = &-\sum_{i = A,B}\Gamma_{i}(S^{z}_{i}S^{z}_{i}\rho-2S^{z}_{i}\rho S^{z}_{i}+\rho S^{z}_{i}S^{z}_{i})\nonumber\\
&&-2\Gamma_{0}(S^{z}_{A}S^{z}_{B}\rho-S^{z}_{B}\rho S^{z}_{A}+\rho S^{z}_{A}S^{z}_{B}-S^{z}_{A}\rho S^{z}_{B}),\nonumber\\
\eea
where $\Gamma_{A} (\Gamma_{B})$ and $2\Gamma_{0}$ are respectively the independent and correlated  dephasing rate of qubit A (B) . The dynamical evolution of this system can then be studied by solving the quantum-Louiville equation (\ref{3}) for $v = 0$ and including the effect of environment by using (\ref{28}). We now consider as earlier that the initial state of the two qubits is defined by the density matrix $\rho$ (\ref{9}). Then the solution of the quantum-Louiville equation for this model of the environment is  given by,
\be
\label{29}
\rho_{11}(t) =  \frac{1}{3}a,\quad \rho_{22}(t) = \frac{1}{3}b,\quad \rho_{33}(t) = \frac{1}{3}c,\nonumber\\
\ee
\be
\label{30}
\rho_{23}(t) = \frac{1}{3}|z|e^{-\left(\Gamma_{A}+\Gamma_{B}-2\Gamma_{0}\right)t}e^{i\chi}\nonumber\\
\ee
\be
\label{31}
\rho_{32}(t) = \rho^{\ast}_{23}(t),\quad \rho_{44}(t) = 1-\rho_{11}(t)-\rho_{22}(t)-\rho_{33}(t)\\
\ee
All other matrix elements of the two qubit density matrix $\rho$ are zero. Now using the solutions of (\ref{31}) it is straight forward to show that, for pure dephasing of the qubits, the \textit{form of matrix in} (\ref{9}) \textit{is preserved for all time}. Note that in such a model the populations do not decay as a result of the interaction with the environment whereas the coherences  like $\rho_{23}(t)$ decay as $\sim \rho_{23}(0)e^{-(\Gamma_{A}+\Gamma_{B}-2\Gamma_{0})t}$ for $\chi = 0$ or mod $\pi$. Let us now study the effect of correlated dephasing of the qubits on the dynamics of entanglement. For the initial conditions $ d = 1-a, b= c = |z| = 1$ and $a \geq 0$ on using (\ref{29}) in (\ref{11}) we get the expression for time dependent concurrence as,
\bea
\label{32}
\tilde{C}_{D}(t) = \frac{2}{3}\left\{e^{-2(\Gamma -\Gamma_{0})t} -\sqrt{a(1-a)}\right \}
\eea
where we have assumed $\Gamma_{A} = \Gamma_{B} = \Gamma$ for simplicity. 
 \begin{figure}[!h]
 \vspace{0.3 in}
\begin{center}
\includegraphics[scale = 0.45]{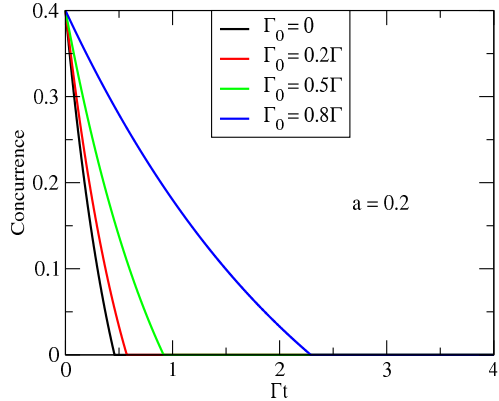}
\caption{(Color online) Time evolution of concurrence for two non-interacting qubits in contact with a purely dephasing environment for initial condition given by $a= 0.2$ and $b = c = |z| = 1$. The effect of correlated dephasing shows up as delay in the onset of ESD.}
\end{center}
\end{figure}
From equation (\ref{32}) it is clearly seen that in a purely dephasing environment entanglement among the qubits is independent of the initial coherence given by $\chi$ and depends only on $a$ and $\Gamma_{0}$. In figure (11) we plot the time dependence of concurrence for $a = 0.2$. We find that the effect of correlated dephasing is manifested in the delay of the onset of ESD. The time for the onset of ESD is given by $t \ge 1/2(\Gamma-\Gamma_{0})\{1/\ln\sqrt{a(1-a)}\}$. From the figure its is clealy visible that with increase in correlated decay $\Gamma_{0}$, the onset of ESD gets delayed further until $\Gamma_{0} = \Gamma$, when concurrence bceomes independent of the dephasing rates and is  given by $C = 2/3[1-\sqrt{a(1-a)}]$. This situation represents a decoherence free subspace where concurrence becomes solely dependent on the value of $a$ \textit{i.e} population of the excited state of the two qubits. Note that this kind of situation has already been tailored to study entanglement in decoherence free subspace \cite{kwait}.\\
\indent{}Let us now include the interaction among the qubits and study how this interaction might influence the entanglement dynamics for this model of the environment. The Hamiltonian of the two qubit system and its coupling to the environment is then given by equations (\ref{1}) and (\ref{28}) respectively. To study the dynamics of entanglement we follow a similar process as described earlier. We use the solution of quantum-Louiville equation derived explicitly in appendix (E) and substitute them in equation (\ref{11}) to calculate the time dependence of concurrence $C$. With the initial conditions $a = 1-d, b = c = |z| = 1$, then we get,
\bea
\label{33}
\tilde{C}_{D}(t) &=& \frac{2}{3}\lbrack e^{-(\Gamma-\Gamma_{0})t}\lbrace e^{-2(\Gamma-\Gamma_{0})t}\cos^{2}\chi\nonumber\\
&+&\sin^{2}\chi(\cos(\Omega^{\prime} t)-\frac{(\Gamma-\Gamma_{0})}{\Omega^{\prime}}\sin(\Omega^{\prime} t))^{2}\rbrace^{1/2}\nonumber\\
&-&\sqrt{a(1-a)}];\\
C(t) & = &\mathsf{Max}\lbrace 0, \tilde{C}_{D}(t)\rbrace
\eea
where $\Omega^{\prime} = \sqrt{4v^{2}-(\Gamma-\Gamma_{0})}$ and we have assumed $\Gamma_{A} = \Gamma_{B}$. One can clearly see the dependence of concurrence on the interaction $v$ among the qubits for $\tilde{C}_{D} > 0$. Note that due to the interaction among the qubits now concurrence  becomes dependent of the initial phase $\chi$.
\begin{figure}[!h]
\vspace{0.35 in}
\begin{center}
\includegraphics[scale = 0.45]{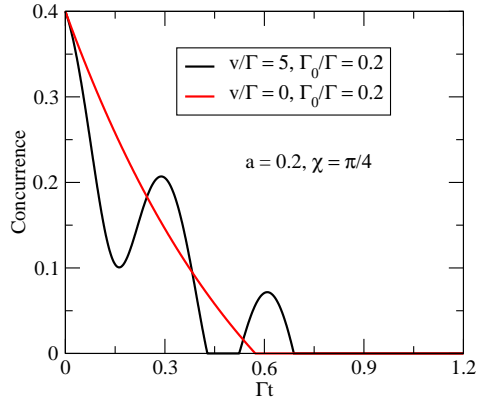}
\caption{(Color online)Time evolution of concurrence for two interacting qubits with interaction strength $v/\Gamma = 5.0$ in contact with a purely correlated dephasing environment and initial condition $a= 0.2., \chi = \pi/4, b = c = |z| = 1.$ The red curve correspond to concurrence of non-interacting qubits. Concurrence is seen to exhibit initial oscillations followed by dark and bright periods with eventual death of entanglement in presence of interaction. The interaction also leads to delayed death of entanglement. Here $\Gamma_{0}$ is the correlated dephasing rate. }
\end{center}
\end{figure}
\begin{figure}[!h]
\vspace{0.35 in}
\begin{center}
\includegraphics[scale = 0.45]{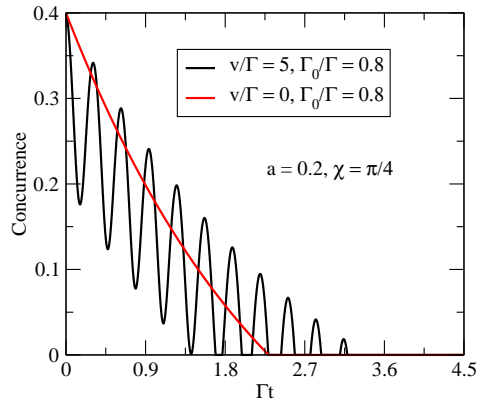}
\caption{(Color online)Time evolution of concurrence for two interacting qubits in contact with a correlated dephasing environment with same parameters as for figure (12) but higher correlated dephasing rates. The effect of higher correlated dephasing manifests itself by increasing the periodicity of dark and bright features in concurrence . Here again we find that dark and bright periods is followed by death of entanglement. }
\end{center}
\end{figure} 
\begin{figure}[!h]
\begin{center}
\includegraphics[scale = 0.45]{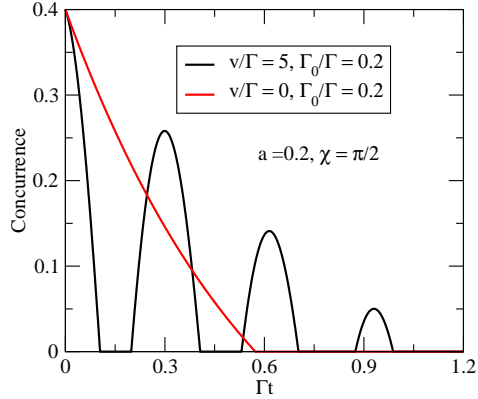}
\caption{(Color online)Time evolution of concurrence for two interacting qubits in contact with a correlated dephasing environment with initial condition $a= 0.2., \chi = \pi/2, b = c = |z| = 1$. Concurrence is seen to be sensitive to initial coherence among the two qubits. It does not exhibit initial oscillations for this value of  $\chi$ but dark and bright periods with eventual death of entanglement in presence of interaction. Here the interaction strength is taken to be $v/\Gamma = 5.0$ }
\end{center}
\end{figure}
To understand the behavior of entanglement in presence of interaction $(v/\gamma = 5.0)$ among the qubits we plot the time dependence of concurrence for different initial phase $\chi$ and correlated dephasing rates $\Gamma_{0}$ in figures (12-14). We consider the case, $a = 0.2$ only to do a comparative study on the behavior of concurrence in presence and absence of inter-qubit interactions. Note that we have already discussed the effect of correlated dephasing on the two qubit entanglement for this value of $a$. Let us now focus on any new feature that arises due to the qubit-qubit interactions. We can see clearly from figure (12) that for $v \neq 0, \chi = \pi/4$ the two qubit concurrence shows a damped oscillatory behavior which leads to dark and bright periods at longer time before eventual death of entanglement. The generation of dark and bright periods is seen to delay the death of entanglement even further in comparison to that induced by correlated dephasing in absence of qubit-qubit interactions. Moreover in figure (13) we see that both the oscillatory behavior as well as dark and bright periods is enhanced with an increase in correlated dephasing rate. When we change the initial phase to $\pi/2$ for $\Gamma_{0} = 0.2\Gamma$ we find (figure 14) no oscillatory behavior in entanglement rather a completely dark and bright periodic feature with eventual delayed death. Thus we see that the onset of dark and bright periods for this kind of environment model is profoundly influenced by the initial coherence of the two qubit system. \\
\indent{}The phenomenon of dark and bright periods in entanglement should have direct consequences for systems like ion traps , quantum dots, the later being currently the forerunner in implementation of quantum logic gates. The interaction between qubits considered in this paper are inherently present in these systems. In quantum dots for example, $\gamma^{-1}\sim$ few ns and one can get a very large range of the parameter $\Gamma^{-1}$ (1-100's of ps) \cite{bori}. Further the interaction strength $v$ can have a range between $1 \mu$ev - $1$ mev depending on gate biasing \cite{atac, kren,beirne, taylor}. An earlier study \cite{gert} reports $\gamma \sim 40-100 \mu$ev and coupling strength of $\sim 100-400 \mu$ev, thereby making $v/\gamma \sim 1- 10$ for quantum dot molecules. Thus experimental parameters are in the range we used for our numerical calculation.
\section{Summary}
In summary we have done a detail study of decoherence effect for non-interacting and interacting initially entangled qubits in contact with different environments at zero temperature . We have shown how the interaction between qubits generates the phenomenon of dark and bright periods in the entanglement dynamics of an intially entangled two qubit system in contact with different environments. We found this feature of dark and bright periods to be generic and occurs for various models of the environment , as an example in a correlated dissipative environment we found the phenomenon of dark and bright periods in entanglement dynamics even though there is no sudden death of entanglement. Moreover for purely dephasing models of the environment we found that the dark and bright periods feature sustains longer and delays the sudden death. We found that there is no sudden death of entanglement for a correlated dissipative environment but rather depending on the initial coherences in the system entanglement can show a substantial slower decay and even the phenomenon of dark and bright periods. For a simple pure dephasing environment as well as for correlated dephasing environment we have shown the existence of sudden death of entanglement. Due to correlated dephasing we found delayed death of entanglement. Further, in the correlated dephasing model we found that the onset of dark and bright periods is sensitive to the initial coherence in the system. The frequency of dark and bright periods was found to depend on the strength of interaction between the qubits as well as on the correlated decay and dephasing rates. As a future perspective it would be interesting to study the effect of qubit-qubit interaction for environments having temperature fluctuations. Further it would also be interesting to extend our study to multi-qubit entanglement. An important class of states that can be treated for this purpose are the GHZ  and W states. Moreover cluster states \cite{clus}  can also be considered as other probable candidates for the study of decoherence and loss of entanglement.\\
This work was supported by NSF grant no CCF-0829860.
\appendix
\section{Solution of the quantum-Louiville equation in the two qubit product basis for two interacting qubits in contact with a dissipative environment.}
\bea
\rho_{11}(t)& = & \rho_{11}(0)e^{-2\gamma t} ,\nonumber\\
\rho_{22}(t) &= &\frac{1}{2}\rho_{22}(0)e^{-\gamma t}(1+\cos(2vt))+\frac{1}{2}\rho_{33}(0)e^{-\gamma t}(1-\cos(2vt))\nonumber\\
&+&\rho_{11}(0)e^{-2\gamma t}(e^{\gamma t}-1)-\frac{i}{2}(\rho_{32}(0)-\rho_{23}(0))e^{-\gamma t}\sin(2vt)\nonumber\\
\rho_{33}(t) &= &\frac{1}{2}\rho_{22}(0)e^{-\gamma t}(1-\cos(2vt))+\frac{1}{2}\rho_{33}(0)e^{-\gamma t}(1+\cos(2vt))\nonumber\\
&+&\rho_{11}(0)e^{-2\gamma t}(e^{\gamma t}-1)+\frac{i}{2}(\rho_{32}(0)-\rho_{23}(0))e^{-\gamma t}\sin(2vt)\nonumber\\
\rho_{12}(t)& = & \rho_{12}(0)e^{-3\gamma t/2}\cos(vt)+i\rho_{13}(0)e^{-3\gamma t/2}\sin(vt)\nonumber\\
\rho_{13}(t)& = & \rho_{13}(0)e^{-3\gamma t/2}\cos(vt)+i\rho_{12}(0)e^{-3\gamma t/2}\sin(vt)\nonumber\\
\rho_{14}(t)& = & \rho_{14}(0)e^{-\gamma t}\nonumber\\
\rho_{23}(t)& = & \frac{i}{2}(\rho_{22}(0)-\rho_{33}(0))\sin(2vt)e^{-\gamma t}+\frac{1}{2}\rho_{23}(0)e^{-\gamma t}(1+\cos(2vt))\nonumber\\
&+&\frac{1}{2}\rho_{32}(0)e^{-\gamma t}(1-\cos(2vt))\nonumber\nonumber
\eea
\bea
\rho_{24}(t) & = & \rho_{24}(0)e^{-\gamma t/2}\cos(vt)-i\rho_{34}(0)e^{-\gamma t/2}\sin(vt)-\rho_{12}(0)\left(\frac{1}{v^{2}+(9/4)\gamma^{2}}\right)[2iv e^{-2\gamma t}\nonumber\\
&+&e^{-\gamma t/2}\{2v\cos(vt)-3i\gamma\sin(vt)\}]-\rho_{13}(0)\left(\frac{1}{v^{2}+(9/4)\gamma^{2}}\right)[3\gamma e^{-2\gamma t}\nonumber\\
&-&e^{-\gamma t/2}\{{2v\sin(vt)+3\gamma\cos(vt)}\}]\nonumber\nonumber
\eea
\bea
\label{4a}
\rho_{34}(t) & = & \rho_{34}(0)e^{-\gamma t/2}\cos(vt)-i\rho_{24}(0)e^{-\gamma t/2}\sin(vt)\nonumber\\
&-&\rho_{12}(0)\left(\frac{1}{v^{2}+(9/4)\gamma^{2}}\right)[3\gamma e^{-2\gamma t}-e^{-\gamma t/2}\{2v\sin(vt)+3\gamma\cos(vt)\}]\nonumber\\
&-&\rho_{13}(0)\left(\frac{1}{v^{2}+(9/4)\gamma^{2}}\right)[2iv e^{-2\gamma t}+e^{-\gamma t/2}\{{2v\cos(vt)-3i\gamma\sin(vt)}\}]\nonumber\\
\eea
and $\rho_{32}(t) = \rho^{\ast}_{23}(t), \rho_{21}(t) = \rho^{\ast}_{12}(t), \rho_{31}(t) = \rho^{\ast}_{13}(t), \rho_{41}(t) = \rho^{\ast}_{14}(t), \rho_{42}(t) = \rho^{\ast}_{24}(t), \rho_{43}(t) = \rho^{\ast}_{34}(t)$, $\rho_{44}(t) = 1-\rho_{11}(t)-\rho_{22}(t)-\rho_{33}(t)$. Note that here we have considered $\gamma_{A} = \gamma_{B} = \gamma$.
\section{Solution of the quantum-Louiville equation in the two qubit product basis for two interacting qubits in contact with a purely dephasing environment. The solutions correspond to the initial matrix $\rho$ defined in equation (\ref{9})}
\bea
\label{70}
\rho_{11}(t)& = & \rho_{11}(0)\quad,\\
\rho_{22}(t) &= &\frac{1}{2}\rho_{22}(0)\left\lbrack1+e^{-(\Gamma_{A}+\Gamma_{B})t/2}\left\lbrace\cos\left(2\Omega t\right)+\frac{(\Gamma_{A}+\Gamma_{B})}{4\Omega}\sin\left(2 \Omega t\right)\right\rbrace\right\rbrack\nonumber\\
&+&\frac{1}{2}\rho_{33}(0)\left\lbrack1-e^{-(\Gamma_{A}+\Gamma_{B})t/2}\left\lbrace\cos\left(2\Omega t\right)+\frac{(\Gamma_{A}+\Gamma_{B})}{4\Omega}\sin\left(2\Omega t\right)\right\rbrace\right\rbrack\nonumber\\
&+&\frac{i[\rho_{23}(0)-\rho_{32}(0)]v e^{-(\Gamma_{A}+\Gamma_{B})t/2}}{2\Omega}\sin\left(2\Omega t\right),\nonumber\\
\\
\rho_{33}(t) &= &\frac{1}{2}\rho_{22}(0)\left\lbrack1-e^{-(\Gamma_{A}+\Gamma_{B})t/2}\left\lbrace\cos\left(2\Omega t\right)+\frac{(\Gamma_{A}+\Gamma_{B})}{4\Omega}\sin\left(2\Omega t\right)\right\rbrace\right\rbrack\nonumber\\
&+&\frac{1}{2}\rho_{33}(0)\left\lbrack1+e^{-(\Gamma_{A}+\Gamma_{B})t/2}\left\lbrace\cos\left(2\Omega t\right)+\frac{(\Gamma_{A}+\Gamma_{B})}{4\Omega}\sin\left(2\Omega t\right)\right\rbrace\right\rbrack\nonumber\\
&-&\frac{i[\rho_{23}(0)-\rho_{32}(0)]v e^{-(\Gamma_{A}+\Gamma_{B})t/2}}{2\Omega}\sin\left(2\Omega t\right),\nonumber\\
\\
\rho_{23}(t) &= &\frac{1}{2}e^{-(\Gamma_{A}+\Gamma_{B})t/2}[\rho_{23}(0)\lbrace e^{-(\Gamma_{A}+\Gamma_{B})t/2}+\cos\left(2\Omega t\right)+\frac{(\Gamma_{A}+\Gamma_{B})}{4\Omega}\sin\left(2\Omega t\right)\rbrace\nonumber\\
&+& \rho_{32}(0)\lbrace e^{-(\Gamma_{A}+\Gamma_{B})t/2}-\cos\left(2\Omega t\right)-\frac{(\Gamma_{A}+\Gamma_{B})}{4\Omega}\sin\left(2\Omega t\right)\rbrace]\nonumber\\
&+&\frac{iv e^{-(\Gamma_{A}+\Gamma_{B})t/2}}{2\Omega}\sin\left(2\Omega t\right)[\rho_{22}(0)-\rho_{33}(0)],\nonumber\\
\\
\Omega & = & \sqrt {v^{2} - \left(\frac{\Gamma_{A}+\Gamma_{B}}{4}\right)^{2}}.
\eea
\be
\rho_{32}(t) = \rho^{\ast}_{23}(t),\quad \rho_{44}(t) = 1-\rho_{11}(t)-\rho_{22}(t)-\rho_{33}(t).
\ee
All other elements of the density matrix $\rho$ defined in the two qubit product basis (\ref{3}) remains zero for all time t.
\section{Solution of the quantum-Louiville equation in the two qubit product basis for two non-interacting qubits in contact with a dissipative environment which results in correlated decay. The solutions correspond to the initial matrix $\rho$ defined in equation (\ref{9})}
\bea
\rho_{11}(t) & = &\rho_{11}(0) e^{-2\gamma t}\quad;\\
\rho_{22}(t) & = & \frac{1}{2}\rho_{22}(0) e^{-\gamma t}(1+\cosh(\Gamma_{12}t))-\frac{1}{2}\rho_{33}(0)e^{-\gamma t}(1-\cosh(\Gamma_{12}t))\nonumber\\
&-&\rho_{11}(0) e^{-2\gamma t}\left(\frac{\gamma^{2}+\Gamma^{2}_{12}}{\gamma^{2}-\Gamma^{2}_{12}}\right)-\frac{1}{2}[\rho_{23}(0)+\rho_{32}(0)] e^{-\gamma t}\sinh(\Gamma_{12}t)\nonumber\\
&+& \frac{1}{2}\rho_{11}(0) e^{-\gamma t}\left[\left(\frac{\gamma+\Gamma_{12}}{\gamma-\Gamma_{12}} \right)e^{\Gamma_{12}t}+\left(\frac{\gamma-\Gamma_{12}}{\gamma+\Gamma_{12}}\right)e^{-\Gamma_{12}t}\right];\nonumber\\
\\
\rho_{33}(t) & = & \frac{1}{2}\rho_{33}(0) e^{-\gamma t}(1+\cosh(\Gamma_{12}t))-\frac{1}{2}\rho_{22}(0) e^{-\gamma t}(1-\cosh(\Gamma_{12}t))\nonumber\\
&-&\rho_{11}(0) e^{-2\gamma t}\left(\frac{\gamma^{2}+\Gamma^{2}_{12}}{\gamma^{2}-\Gamma^{2}_{12}}\right)-\frac{1}{2}[\rho_{23}(0)+\rho_{32}(0)] e^{-\gamma t}\sinh(\Gamma_{12}t)\nonumber\\
&+&\frac{1}{2}\rho_{11}(0) e^{-\gamma t}\left[\left(\frac{\gamma+\Gamma_{12}}{\gamma-\Gamma_{12}} \right)e^{\Gamma_{12}t}+\left(\frac{\gamma-\Gamma_{12}}{\gamma+\Gamma_{12}}\right)e^{-\Gamma_{12}t}\right];\nonumber\\
\\
\rho_{23}(t) & = & \frac{1}{2}\rho_{23}(0)e^{-\gamma t}(1+\cosh(\Gamma_{12}t))-\frac{1}{2}\rho_{32}(0)e^{-\gamma t}(1-\cosh(\Gamma_{12}t))\nonumber\\
&-&\frac{1}{2}[\rho_{22}(0)+\rho_{33}(0)]e^{-\gamma t}\sinh(\Gamma_{12}t)-\rho_{11}(0) e^{-2\gamma t}\left(\frac{2\gamma\Gamma_{12}}{\gamma^{2}-\Gamma^{2}_{12}}\right)\nonumber\\
&+&\frac{1}{2}\rho_{11}(0) e^{-\gamma t}\left[\left(\frac{\gamma+\Gamma_{12}}{\gamma-\Gamma_{12}} \right)e^{\Gamma_{12}t}+\left(\frac{\gamma-\Gamma_{12}}{\gamma+\Gamma_{12}}\right)e^{-\Gamma_{12}t}\right];\nonumber\\
\eea
\be
\rho_{32}(t) = \rho_{23}^{\ast}(t), \quad \rho_{44}(t) = 1-\rho_{11}(t)-\rho_{22}(t)-\rho_{33}(t).
\ee
All other elements of the density matrix $\rho$ defined in the two qubit product basis (\ref{3}) remains zero for all time t.
\section{Solution of the quantum-Louiville equation in the two qubit product basis for two interacting qubits in contact with a dissipative environment which results in correlated decay. The solutions correspond to the initial matrix $\rho$ defined in equation (\ref{9})}
\bea
\rho_{11}(t) & = &\rho_{11}(0) e^{-2\gamma t}\quad;\\
\rho_{22}(t) & = & \frac{1}{2}\rho_{22}(0) e^{-\gamma t}(\cos(2vt)+\cosh(\Gamma_{12}t))-\frac{1}{2}\rho_{33}(0) e^{-\gamma t}(\cos(2vt)-\cosh(\Gamma_{12}t))\nonumber\\
&-&\rho_{11}(0) e^{-2\gamma t}\left(\frac{\gamma^{2}+\Gamma^{2}_{12}}{\gamma^{2}-\Gamma^{2}_{12}}\right)+\frac{1}{2}\rho_{11}(0)e^{-\gamma t}\left[\left(\frac{\gamma+\Gamma_{12}}{\gamma-\Gamma_{12}} \right)e^{\Gamma_{12}t}+\left(\frac{\gamma-\Gamma_{12}}{\gamma+\Gamma_{12}}\right)e^{-\Gamma_{12}t}\right]\nonumber\\
&-&\frac{1}{2}[\rho_{23}(0)+\rho_{32}(0)]e^{-\gamma t}\sinh(\Gamma_{12}t)+\frac{1}{2}[\rho_{23}(0)-\rho_{32}(0)]e^{-\gamma t}\sin(2vt);\nonumber\\
\\
\rho_{33}(t) & = & \frac{1}{2}\rho_{33}(0) e^{-\gamma t} (\cos(2vt)+\cosh(\Gamma_{12}t))-\frac{1}{2}\rho_{22}(0) e^{\gamma t}(\cos(2vt)-\cosh(\Gamma_{12}t))\nonumber\\
&-&\rho_{11}(0) e^{-2\gamma t}\left(\frac{\gamma^{2}+\Gamma^{2}_{12}}{\gamma^{2}-\Gamma^{2}_{12}}\right)+\frac{1}{2}\rho_{11}(0)e^{-\gamma t}\left[\left(\frac{\gamma+\Gamma_{12}}{\gamma-\Gamma_{12}} \right)e^{\Gamma_{12}t}+\left(\frac{\gamma-\Gamma_{12}}{\gamma+\Gamma_{12}}\right)e^{-\Gamma_{12}t}\right]\nonumber\\
&-&\frac{1}{2}[\rho_{23}(0)+\rho_{32}(0)]e^{-\gamma t}\sinh(\Gamma_{12}t)-\frac{1}{2}[\rho_{23}(0)-\rho_{32}(0)]e^{-\gamma t}\sin(2vt);\nonumber\\
\\
\rho_{23}(t) & = & \frac{1}{2}\rho_{23}(0)e^{-\gamma t}(\cos(2vt)+\cosh(\Gamma_{12}t))-\frac{1}{2}\rho_{32}(0)e^{-\gamma t}(\cos(2vt)-\cosh(\Gamma_{12}t))\nonumber\\
&-&\rho_{11}(0)e^{-2\gamma t}\left(\frac{2\gamma\Gamma_{12}}{\gamma^{2}-\Gamma^{2}_{12}}\right)+\frac{1}{2}\rho_{11}(0) e^{-\gamma t}\left[\left(\frac{\gamma+\Gamma_{12}}{\gamma-\Gamma_{12}} \right)e^{\Gamma_{12}t}+\left(\frac{\gamma-\Gamma_{12}}{\gamma+\Gamma_{12}}\right)e^{-\Gamma_{12}t}\right]\nonumber\\
&-&\frac{1}{6}[\rho_{22}(0)+\rho_{33}(0)]e^{-\gamma t}\sinh(\Gamma_{12}t)+\frac{1}{2}e^{-\gamma t}[\rho_{22}(0)-\rho_{33}(0)]i\sin(2vt);\nonumber\\
\eea
\be
\rho_{32}(t) = \rho_{23}^{\ast}(t), \quad \rho_{44}(t) = 1-\rho_{11}(t)-\rho_{22}(t)-\rho_{33}(t).
\ee
All other elements of the density matrix $\rho$ defined in the two qubit product basis (\ref{3}) remains zero for all time t.
\section{Solution of the quantum-Louiville equation in the two qubit product basis for two interacting qubits in contact with a purely dephasing environment. The solutions correspond to the initial matrix $\rho$ defined in equation (\ref{9})}
\bea
\label{71}
\rho_{11}(t)& = & \rho_{11}(0)\quad,\\
\rho_{22}(t) &= &\frac{1}{2}\rho_{22}(0)\left[1+e^{-(\Gamma_{A}+\Gamma_{B}-2\Gamma_{0})t/2}\cos\left(2\Omega^{\prime} t\right)+\frac{(\Gamma_{A}+\Gamma_{B}-2\Gamma_{0})}{4\Omega^{\prime}}\sin\left(2\Omega^{\prime}t\right)\right\rbrack\nonumber\\
&+& \frac{1}{2}\rho_{33}(0)\left[1-e^{-(\Gamma_{A}+\Gamma_{B}-2\Gamma_{0})t/2}\cos\left(2\Omega^{\prime}t\right)+\frac{(\Gamma_{A}+\Gamma_{B}-2\Gamma_{0})}{4\Omega^{\prime}}\sin\left(2\Omega^{\prime}t\right)\right]\nonumber\\
&+&\frac{i[\rho_{23}(0)-\rho_{32}(0)]v e^{-(\Gamma_{A}+\Gamma_{B}-2\Gamma_{0})t/2}}{2\Omega^{\prime}}\sin\left(2\Omega^{\prime}t\right)\quad,\\
\rho_{33}(t) &= &\frac{1}{2}\rho_{22}(0)\left[1-e^{-(\Gamma_{A}+\Gamma_{B}-2\Gamma_{0})t/2}\cos\left(2\Omega^{\prime} t\right)+\frac{(\Gamma_{A}+\Gamma_{B}-2\Gamma_{0})}{4\Omega^{\prime}}\sin\left(2\Omega^{\prime}t\right)\right\rbrack\nonumber\\
&+& \frac{1}{2}\rho_{33}(0)\left[1+e^{-(\Gamma_{A}+\Gamma_{B}-2\Gamma_{0})t/2}\cos\left(2\Omega^{\prime}t\right)+\frac{(\Gamma_{A}+\Gamma_{B}-2\Gamma_{0})}{4\Omega^{\prime}}\sin\left(2\Omega^{\prime}t\right)\right]\nonumber\\
&-&\frac{i[\rho_{23}(0)-\rho_{32}(0)]v e^{-(\Gamma_{A}+\Gamma_{B}-2\Gamma_{0})t/2}}{2\Omega^{\prime}}\sin\left(2\Omega^{\prime}t\right)\quad,\\
\rho_{23}(t) &= &\frac{1}{2}e^{-(\Gamma_{A}+\Gamma_{B}-2\Gamma_{0})t/2}[\rho_{23}(0)\lbrace e^{-(\Gamma_{A}+\Gamma_{B}-2\Gamma_{0})t/2}\nonumber\\
&+&\cos\left(2\Omega^{\prime}t\right)+\frac{(\Gamma_{A}+\Gamma_{B}-2\Gamma_{0})}{4\Omega^{\prime}}\sin\left(2\Omega^{\prime}t\right)\rbrace\nonumber\\
&+&\rho_{32}(0)\lbrace e^{-(\Gamma_{A}+\Gamma_{B}-2\Gamma_{0})t/2}-\cos\left(2\Omega^{\prime}t\right)-\frac{(\Gamma_{A}+\Gamma_{B}-2\Gamma_{0})}{4\Omega^{\prime}}\sin\left(2\Omega^{\prime}t\right)\rbrace]\nonumber\\
&+&\frac{iv e^{-(\Gamma_{A}+\Gamma_{B}-2\Gamma_{0})t/2}}{2\Omega^{\prime}}\sin\left(2\Omega^{\prime}t\right)[\rho_{22}(0)-\rho_{33}(0)]\quad,\nonumber\\
\\
\nonumber\\
\Omega^{\prime} & = & \sqrt{v^{2}-\left(\frac{\Gamma_{A}+\Gamma_{B}-2\Gamma_{0}}{4} \right)^{2}}
\eea
\\
\be
\rho_{32}(t) = \rho^{\ast}_{23}(t),\quad \rho_{44}(t) = 1-\rho_{11}(t)-\rho_{22}(t)-\rho_{33}(t).
\ee
All other elements of the density matrix $\rho$ defined in the two qubit product basis (\ref{3}) remains zero for all time t.

\end{document}